\documentclass[final,5p,times,twocolumn]{elsarticle}

\usepackage[utf8]{inputenc}
\usepackage{amsmath}
\usepackage{graphicx}
\usepackage{subfig}
\usepackage{siunitx}
\DeclareSIUnit\angstrom{\text{\AA}}
\usepackage{textcomp}
\usepackage{gensymb}
\usepackage{microtype}
\usepackage{amsfonts}
\usepackage{multirow}
\usepackage{lmodern}
\usepackage{caption}
\usepackage[colorlinks=true,linkcolor=cyan,citecolor=cyan,urlcolor=cyan]{hyperref}

\makeatletter
\def\ps@pprintTitle{%
  \let\@oddhead\@empty
  \let\@evenhead\@empty
  \def\@oddfoot{}%
\let\@evenfoot\@oddfoot}
\makeatother

\begin{document}

\twocolumn[
  \begin{@twocolumnfalse}
    \begin{center}
      {\Large\bfseries Maskless Electron Beam–Induced Etching of Diamond in Air: A Secondary Electron–Driven Mechanism\par}
      \vspace{1.0em}
      {\normalsize Duc-Duy Tran$^{1,2,3}$, Cedric Mannequin$^{3,4}$, Fabrice Donatini$^{1}$, Masahiro Sasaki$^{2}$, Etienne Gheeraert$^{1,2,3}$\par}
      \vspace{0.8em}
      {\small\itshape
        $^1$Univ. Grenoble Alpes, CNRS, Grenoble INP, Institut Neel, Grenoble 38000, France\\
        $^2$Institute of Applied Physics, Faculty of Pure and Applied Sciences, University of Tsukuba, Tsukuba 305-8573, Japan\\
        $^3$Japanese-French Laboratory for Semiconductor Physics and Technology J-FAST, CNRS, Université Grenoble Alpes, Grenoble INP, University of Tsukuba, Japan\\
        $^4$CNRS-Nantes Université-Institut des Matériaux de Nantes Jean Rouxel, IMN, F-44000 Nantes, France
      \par}
    \end{center}

    \vspace{0.8em}
    \hrule
    \vspace{0.8em}

    {\Large\bfseries Abstract\par}
    \vspace{0.5em}

    We report a direct, maskless electron beam–induced etching (EBIE) process for diamond in air, enabling high-precision patterning without lithography or plasma processing.

    Through a comprehensive analysis of electron–gas, electron–diamond, and gas–surface interactions in the SEM environment, we demonstrate that etching is predominantly governed by low-energy secondary electrons, which drive gas dissociation and radical generation.

    The resulting oxygen- and nitrogen-based radicals chemisorb on the diamond surface, form volatile carbon-containing species, and desorb under continued electron irradiation, enabling controlled material removal.

    The process exhibits two distinct regimes: a molecule-limited regime governed by gas flux and an electron-limited regime controlled by current density.

    Etch depths up to 212\,nm and lateral resolution down to 200\,nm are achieved. Time-dependent anisotropy is observed, with (100) surfaces transitioning to (111)-faceted morphologies, enhancing etch yield.

    These results establish a general secondary electron–driven mechanism for EBIE in gas environments, providing a maskless, damage-free nanofabrication route for diamond semiconductor and other chemically inert materials.

    \vspace{0.6em}
    \textit{Keywords:} Diamond, EBIE, Electron beam induced etching, Air, Secondary electrons, Anisotropic etching

    \vspace{0.8em}
    \hrule
    \vspace{1.0em}
  \end{@twocolumnfalse}
]

\section{Introduction}

Diamond possesses exceptional physical and electronic properties, including the highest known thermal conductivity (22 W·cm$^{-1}$·K$^{-1}$ at room temperature \cite{umezawa2018recent}), ultrahigh hole mobilities (up to $10^6$ cm$^2$·V$^{-1}$·s$^{-1}$ \cite{portier2023carrier}), a high critical electric field (9.5 MV·cm$^{-1}$ \cite{volpe2010high}), and an ultra-wide bandgap of 5.47 eV \cite{umezawa2018recent}. These characteristics make diamond a highly attractive material for advanced technologies such as high-power electronics \cite{donato2019diamond}, alpha-particle radiation detectors \cite{pomorski2013super}, and photonic and quantum devices \cite{mi2020integrated}.

Despite its technological promise, diamond remains difficult to process due to its extreme hardness and chemical inertness. Current top-down fabrication methods rely primarily on plasma-based reactive ion etching (RIE) \cite{toros2020reactive}, a complex, multi-step process involving mask deposition, lithography, etching, and subsequent mask removal. This approach is not only time-consuming and resource-intensive but also introduces surface and subsurface damage through high-energy ion bombardment, which can degrade device performance \cite{kawabata2004xps}.

In this work, we develop electron beam–induced etching EBIE as a rapid, maskless, and high-precision technique for fabricating diamond micro- and nanoelectronic structures. The etching is carried out in SEM chamber, where the diamond surface is exposed to an electron beam in the presence of an air ambient. This approach avoids the surface and subsurface damage associated with ion bombardment and enables direct etching without the need for lithographic masks, allowing precise control over feature shape and position. The ability to preserve the pristine surface quality is particularly attractive for photonic and quantum devices, where structural damage and the formation of dead layers must be strictly avoided to maintain coherence, low noise, and quantum stability. EBIE of diamond has previously been demonstrated in hydrogen and oxygen atmospheres~\cite{taniguchi1994electron, taniguchi1997electron}, with additional studies exploring nitrogen, water vapor, and air as alternative etching gases~\cite{bishop2018deterministic, niitsuma2006nanoprocessing}.

At room temperature and standard conditions (1~bar, 20\,\textdegree{}C), diamond is chemically inert in most environments~\cite{pierson2012handbook}. In vacuum, and for electron energies below 30\,keV, physical sputtering is negligible~\cite{utke2012nanofabrication}. Thus, enabling EBIE requires two key steps: (i) chemisorption of dissociated or activated gas species onto the diamond surface, and (ii) desorption of volatile reaction products, typically driven by primary electron (PE) beam energy~\cite{taniguchi1994electron}.

In the SEM chamber, incident primary electrons PEs partially scatter with the surrounding air between the electron gun and the diamond surface. Due to their high energy, most PEs reach the sample surface and interact with air molecules near the surface, as well as with the diamond atoms.

Collisions between PEs and diamond can involve inelastic scattering with atomic electrons or elastic scattering with carbon nuclei. In the case of elastic scattering, the PE may be re-emitted from the surface with minimal energy loss; this re-emitted electron is referred to as a backscattered electron (BSE), and typically retains an energy close to that of the incident PE. In contrast, inelastic scattering may lead to the emission of electrons from the surface if the transferred energy exceeds the work function of the diamond. If the emitted electron has an energy below 50\,eV, it is classified as a secondary electron (SE)~\cite{brundle1992encyclopedia}. Consequently, the re-emitted BSEs and SEs can further scatter with gas molecules located on or near the diamond surface.

Collisions between electrons and gas molecules lead to the generation of reactive species, such as radicals and ions. These electron–gas interactions can occur through elastic scattering, vibrational excitation, electronic excitation, or dissociation scattering \cite{utke2012nanofabrication}. While the first three mechanisms alter the physical state of the gas molecules, dissociation scattering induces a chemical change by breaking molecular bonds \cite{utke2012nanofabrication}. The resulting dissociation products may adsorb on the diamond surface through chemisorption. If these chemisorption product are volatile, they can subsequently desorb from the surface, resulting in diamond etching. Each of these processes is highly energy-dependent.

In this work, we present a comprehensive investigation of the mechanisms underlying electron beam-induced etching EBIE of diamond. Our study examines the entire sequence of events—from the emission of electrons from the SEM gun to the final removal of carbon atoms from the diamond surface. This includes detailed analysis of electron–gas, electron–diamond, and gas–diamond interactions occurring within the SEM chamber. We systematically explore how the etching behavior depends on key parameters: primary electron PE energy, the spatial positioning between the gas injection nozzle and the electron beam, electron current density, electron dwell time and PE exposure time. These dependencies are used to elucidate the sequential etching mechanism and to identify optimized conditions for controlled, maskless diamond etching.

\section{Materials and Methods}

A high-pressure high-temperature (HPHT), (100)-oriented single-crystal diamond was used. The sample was lightly doped with nitrogen to prevent surface charging during electron beam exposure. Prior to etching, the sample was sequentially cleaned with acetone, ethanol, and deionized water, followed by immersion in a hot triacid solution ($\mathrm{HClO_4:HNO_3:H_2SO_4}$ in a molar ratio of 1:3:4) at 250\,\textdegree{}C for 30 minutes to remove organic and inorganic surface contaminants.

\begin{figure}[!t]
  \centering
  \includegraphics[width=\columnwidth]{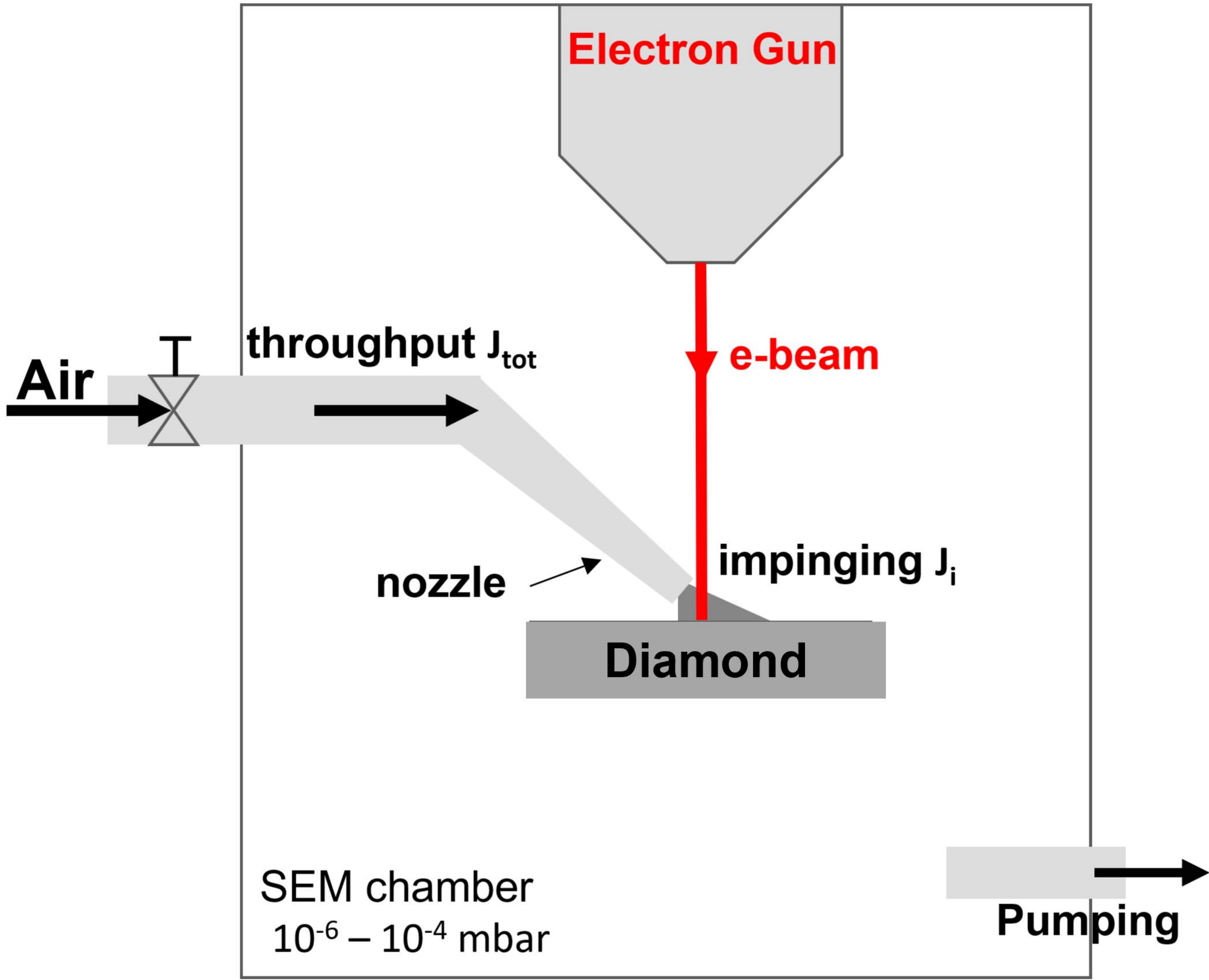}

  \caption{Schematic of the EBIE setup in an SEM chamber. Air is introduced with a total throughput $J_{\mathrm{tot}}$ and directed via a nozzle to produce a local impinging flux $J_i$ toward the diamond surface. An electron beam irradiates the surface under controlled chamber pressure ($10^{-6}$--$10^{-4}$\,mbar), maintained by continuous pumping.}

  \label{fig:SetUp}
\end{figure}

The EBIE process was carried out in an FEI Quanta~200 environmental SEM chamber equipped with a gas injection system, as illustrated in Figure~\ref{fig:SetUp}. The electron beam was generated by a thermionic electron gun operating at an acceleration voltage ranging from 200\,V to 30\,kV. The beam current, calibrated using a Faraday cup, was varied from several hundred pA  to 6\,nA.

The gas injection system consisted of a valve connected to ambient air and a nozzle mounted inside the SEM chamber. The nozzle was positioned at an angle of 30\textdegree{} relative to the diamond surface to direct the air flow toward the irradiated region. The inlet provided atmospheric air composed primarily of nitrogen ($\sim78$\%), oxygen ($\sim21$\%), and water vapor ($\sim1$--3\%)~\cite{gevantman2006crc}. The SEM chamber pressure, corresponding to the nozzle outlet pressure, was maintained in the $10^{-6}$ to $10^{-4}$\,mbar range using continuous pumping with a turbo-molecular pump.

After etching, the diamond surface was imaged using the SEM in image mode at acceleration voltages between 10\,kV and 30\,kV. Surface roughness, etch depth, and etched volume were measured using a Bruker ContourGT optical profilometer.

Diamond etching was studied under varying conditions: electron energy (1--10\,keV), lateral distance between the gas nozzle and the electron beam (0--836\,$\mu$m), electron current density (0--250\,pA/$\mu$m$^{-2}$), gas pressure ($2$--$206 \times 10^{-6}$\,mbar), and exposure time (0--30 minutes).

\section{Result and  Discussion}

\subsection{Electron energy dependence }
\label{sect:Energy_Dependence}

\begin{figure}[!t]
  \centering
  \includegraphics[width=\columnwidth]{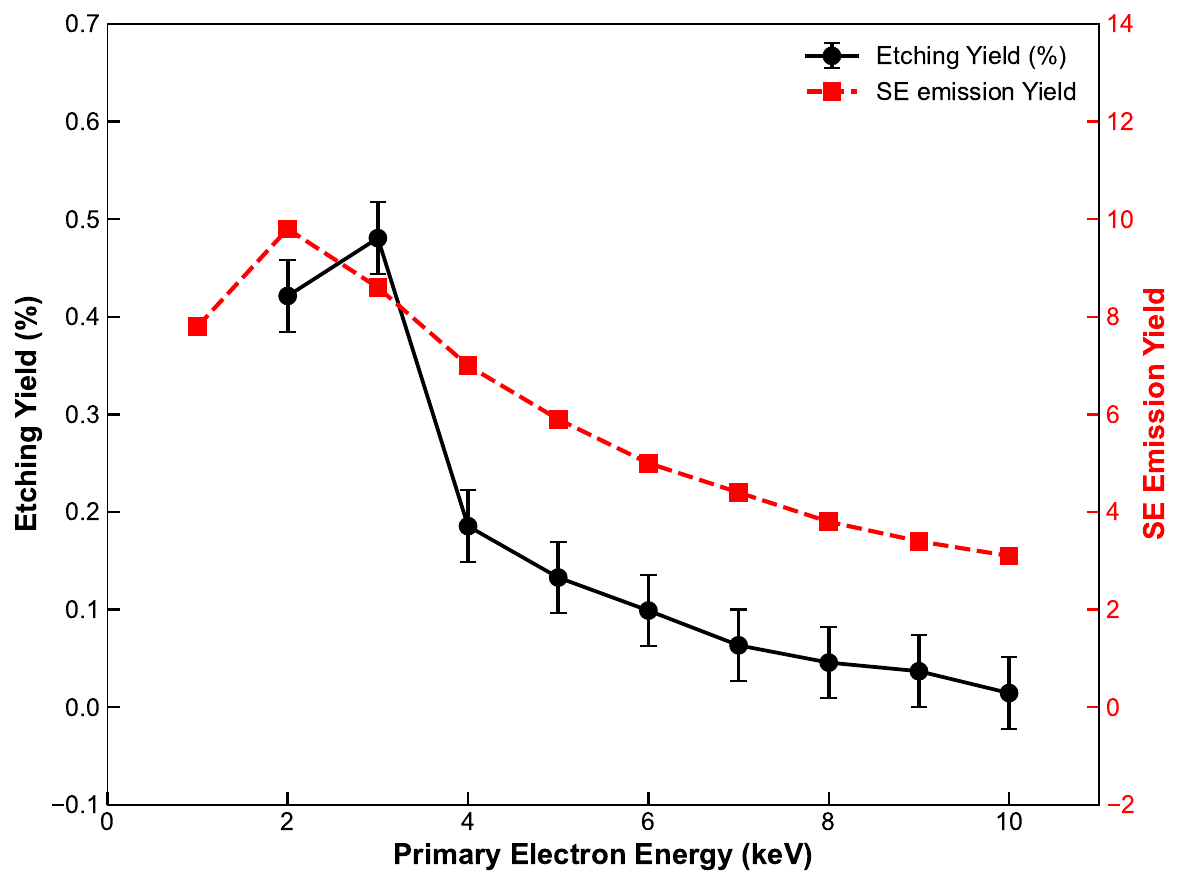}

  \caption{Dependence of etching yield (black circles) and secondary electron SE emission yield (red squares) on primary electron energy.  Etching conditions: $2.5 \times 2.5~\mu\text{m}^2$ square scanning area, $10^{-4}$~mbar air, 30~min exposure. Error bars on the etching yield reflect variations associated with the surface roughness of unetched diamond. Secondary electron yield data are reproduced from Ref.~\cite{ascarelli2001secondary}. \textbf{The observed correlation between SE yield and etching efficiency highlights the influence of secondary electrons in the EBIE process.}}

  \label{fig:Energy_Dependence}
\end{figure}

The dependence of diamond etching yield on primary electron - PE energy in the 2--10\,keV range is presented in Figure~\ref{fig:Energy_Dependence}, where dot symbols represent experimental etching yield and star symbols indicate secondary electron SE emission yield values extracted from Ref.~\cite{ascarelli2001secondary}. The etching yield, $\mu$, is defined as the number of carbon atoms removed ($N_C$) per incident PE, and is calculated as:

\begin{equation}
  \mu = \frac{N_C}{N_e} = \frac{\rho_N V_e q}{I_{\mathrm{PE}} t}
\end{equation}

where $N_e$ is the total number of incident PEs, $\rho_N$ is the carbon atomic number density in diamond (in $\mathrm{m}^{-3}$), $V_e$ is the etched volume measured by optical profilometry (in $\mathrm{m}^3$), $q$ is the elementary charge ($1.6 \times 10^{-19}$\,C), $I_{\mathrm{PE}}$ is the PE current (A), and $t$ is the total etching time (s).

The etching yield increases with PE energy in the 2--3\,keV range, reaching a maximum of approximately 0.5\% at 3\,keV, and subsequently decreases with increasing energy. At 10\,keV, no measurable etching is observed by either SEM imaging or optical profilometry.

The SE emission yield, defined as the number of secondary electrons emitted per incident primary electron, exhibits a strong dependence on PE energy. In the 1--10\,keV range, the SE yield reaches a maximum of approximately 10.5 at 1\,keV and decreases rapidly to around 2.5 at 10\,keV~\cite{ascarelli2001secondary}. The qualitative similarity between the SE emission curve and the measured etching yield suggests that the etching process is predominantly driven by interactions between SEs and gas molecules. These interactions are responsible for generating reactive species that subsequently enable carbon removal from the diamond surface.

\begin{table*}[!t]
  \centering
  \small
  \caption{Maximum dissociation cross sections for electron collisions with air molecules, including nitrogen, oxygen, and water vapor.}
  \label{tab:Gas_Cross_Section}
  \resizebox{\textwidth}{!}{%
    \begin{tabular}{llllrr}
      \hline
      \textbf{Molecule [Ref]} & \textbf{\% in Air} & \textbf{Type} & \textbf{Product} & \textbf{Energy (eV)} & \textbf{Maximum Cross Section ($10^{-18}$ cm$^{-2}$)} \\ \hline
      \multirow{3}{*}{$N_2$ \cite{itikawa2006cross}} & \multirow{3}{*}{78} & Neutral Dissociation & $N^{\cdot}$ & 60 & 123 \\
      & & Ionization Dissociation & $N^{+}$ & 120 & 66 \\
      & & Ionization Dissociation & $N^{++}$ & 200 & 15 \\ \hline
      \multirow{4}{*}{$O_2$ \cite{itikawa2009cross}} & \multirow{4}{*}{21} & Neutral Dissociation & $O^{\cdot}$ & 33.5 & 66 \\
      & & Attachment Dissociation & $O^{-}$ & 6.5 & 1.4 \\
      & & Ionization Dissociation & $O^{+}$ & 138 & 91 \\
      & & Ionization Dissociation & $O^{++}$ & 273 & 2.1 \\ \hline
      \multirow{9}{*}{$H_2O$ \cite{itikawa2005cross}} & \multirow{9}{*}{1--3} & Neutral Dissociation & $O^{\cdot}$ & 106 & 1.5 \\
      & & Neutral Dissociation & $OH^{\cdot}$ & 75 & 210 \\
      & & Attachment Dissociation & $O^{-}$ & 8.44 & 0.316 \\
      & & Attachment Dissociation & $OH^{-}$ & 6.4 & 0.116 \\
      & & Attachment Dissociation & $H^{-}$ & 6.4 & 6.37 \\
      & & Ionization Dissociation & $OH^{+}$ & 100 & 41.8 \\
      & & Ionization Dissociation & $O^{+}$ & 125 & 7.63 \\
      & & Ionization Dissociation & $O^{++}$ & 250 & 0.195 \\
      & & Ionization Dissociation & $H^{+}$ & 175 & 36.6 \\ \hline
    \end{tabular}%
  }
\end{table*}

The probability of gas-electron interactions is governed by the electron scattering cross section, which quantifies the efficiency of collision-induced processes. For the major air constituents—nitrogen, oxygen, and water vapor—the dissociative scattering cross sections peak in the electron energy range of 10–150\,eV and decline sharply at higher energies~\cite{itikawa2006cross, itikawa2009cross, itikawa2005cross}. As SEs typically have energies below 50\,eV, they fall within this optimal dissociation range. In contrast, PEs and BSEs possess significantly higher energies, making their contribution to gas dissociation comparatively minor. Consequently, reactive species are predominantly generated via SE–gas collisions, which play a central role in the EBIE mechanism.

Most of the electrons re-emitted from diamond following PE impact are SEs, with energies below 50\,eV and largely independent of the incident PE energy~\cite{ascarelli2001secondary}. In contrast, the SE emission yield—the number of SEs emitted per PE—depends strongly on the PE energy~\cite{ascarelli2001secondary}. As SE–gas interactions are primarily responsible for generating reactive species, the etching yield is directly correlated with the SE emission yield. Accordingly, the PE energy dependence of the etching yield closely follows that of the SE yield. Notably, the maximum etching yield occurs at a PE energy of 3\,keV, whereas the SE emission yield peaks near 1\,keV. This slight offset may be attributed to material differences: our measurements were performed on lightly doped single-crystal diamond, while the SE data in Ref.~\cite{ascarelli2001secondary} were obtained from polycrystalline diamond.

Table~\ref{tab:Gas_Cross_Section} lists the maximum dissociation cross sections for electron collisions with nitrogen, oxygen, and water vapor molecules~\cite{itikawa2006cross, itikawa2009cross, itikawa2005cross}. The contributions from minor air components such as Ar, CO$_2$, and CH$_4$ are negligible due to their low concentrations (less than 1\%). Notably, the dissociation of nitrogen, oxygen, and water vapor produces reactive radicals—N$^{\cdot}$, O$^{\cdot}$, and OH$^{\cdot}$—with maximum cross sections in the 33–75\,eV energy range. This overlaps well with the typical energy distribution of SEs (below 50\,eV), indicating that these radicals are preferentially generated during EBIE.

Among these species, O$^{\cdot}$ and N$^{\cdot}$ are the primary contributors to etching, as they readily form volatile compounds—CO and CN—upon chemisorption on the diamond surface. In contrast, OH$^{\cdot}$ does not form volatile carbon-containing products under the present conditions and therefore plays a minimal role in material removal. These results suggest that SE-induced dissociation of O$_2$ and N$_2$ is the primary mechanism for radical generation during air-based EBIE.

\subsection{Influence of beam position }

\begin{figure*}[!t]
  \centering
  \subfloat[]{\includegraphics[width=0.49\textwidth]{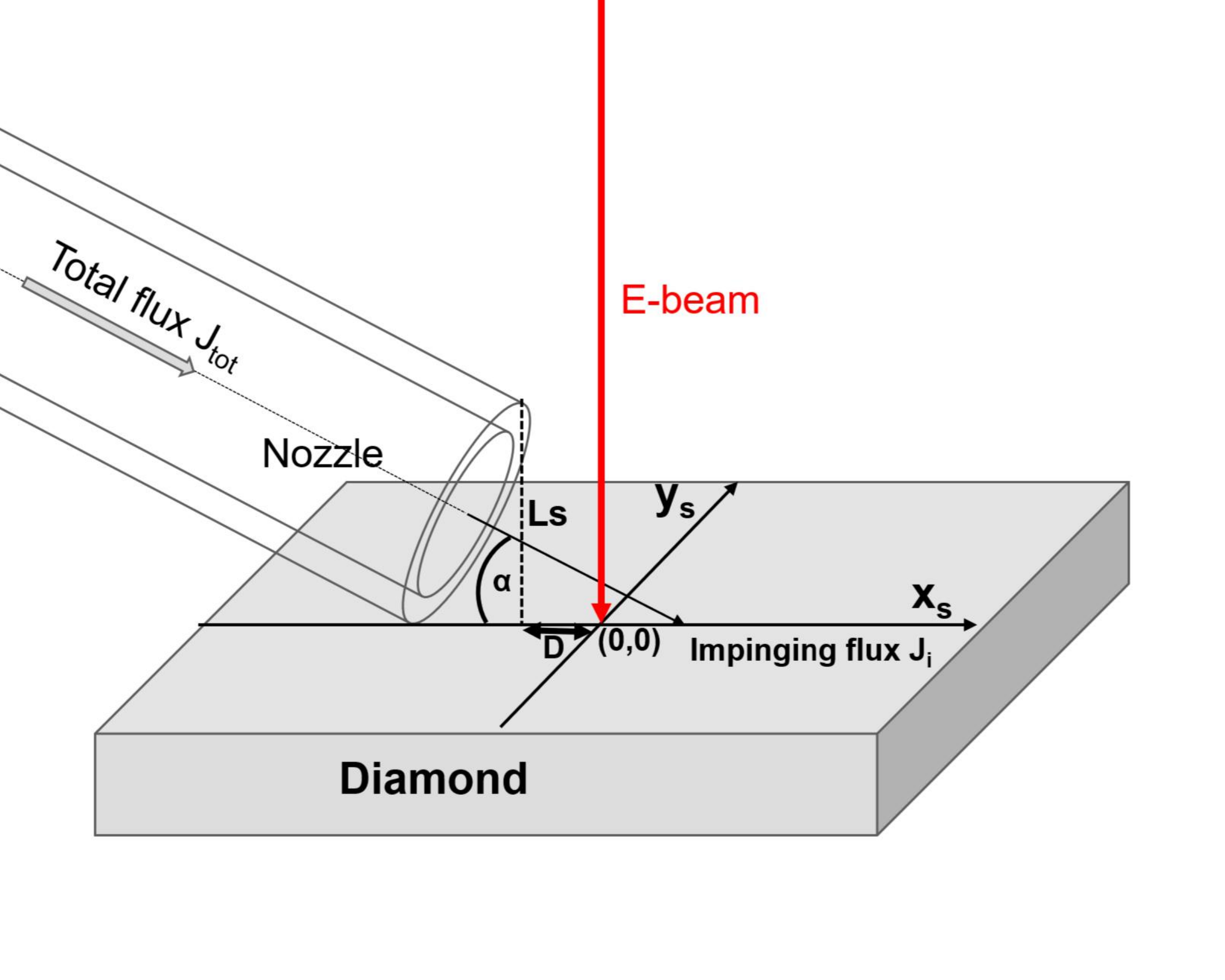}}\hfill
  \subfloat[]{\includegraphics[width=0.49\textwidth]{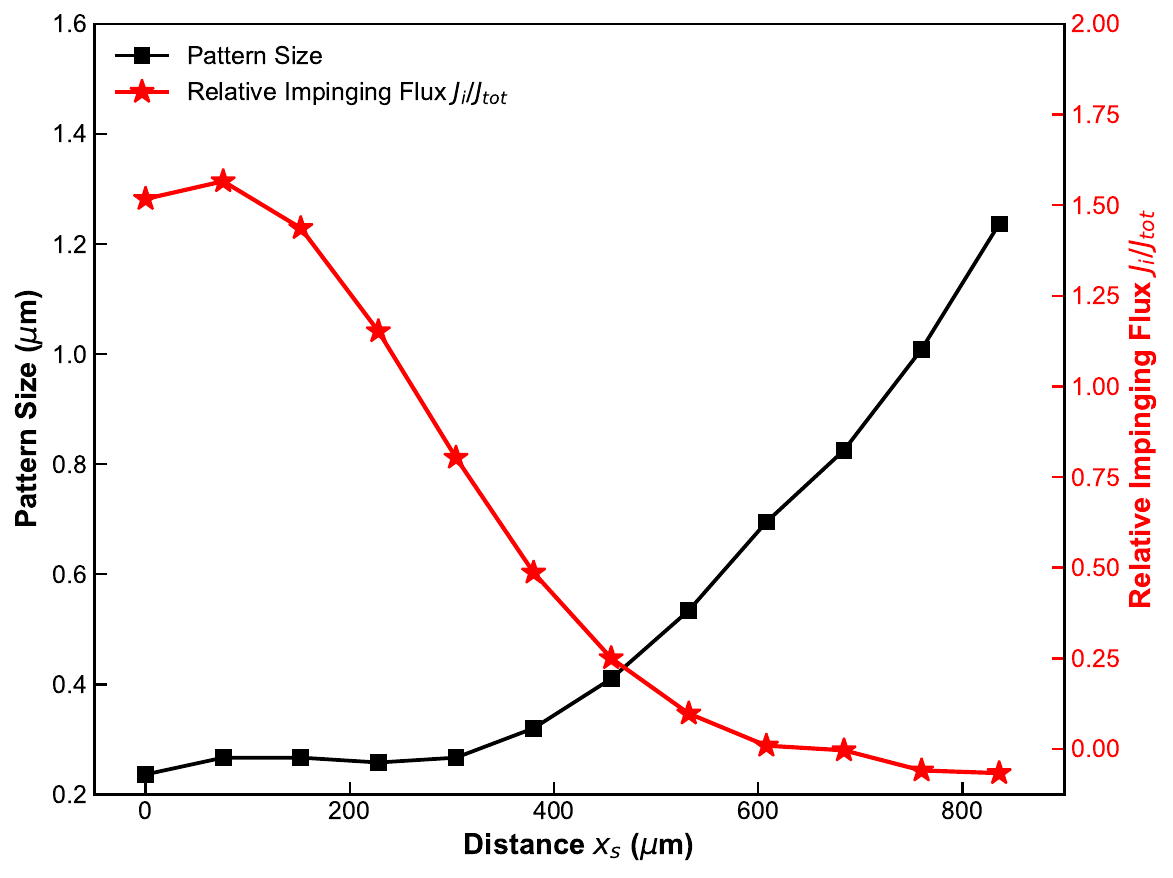}}
  \caption{Effect of nozzle-to-beam distance $x_s$ on etching pattern size.
    (a) Schematic of the EBIE setup with the nozzle positioned at a 30$^\circ$ angle relative to the diamond surface. The origin $x_s = 0$ corresponds to a distance $D = 424\,\mu\mathrm{m}$ from the nozzle projection onto the sample plane.
  (b) Dependence of pattern size and relative impinging flux on $x_s$. Flux values are obtained from geometric simulation under free molecular flow conditions. Etching conditions: spot scanning, 5\,keV electron energy, 0.935\,pA current, $10^{-4}$\,mbar air, 2\,min exposure. \textbf{The strong correlation between pattern size and impinging flux highlights the spatial dependence of EBIE under these conditions.}}
  \label{fig:Distance_Dependence}
\end{figure*}

\begin{figure}[!t]
  \centering
  \includegraphics[width=\columnwidth]{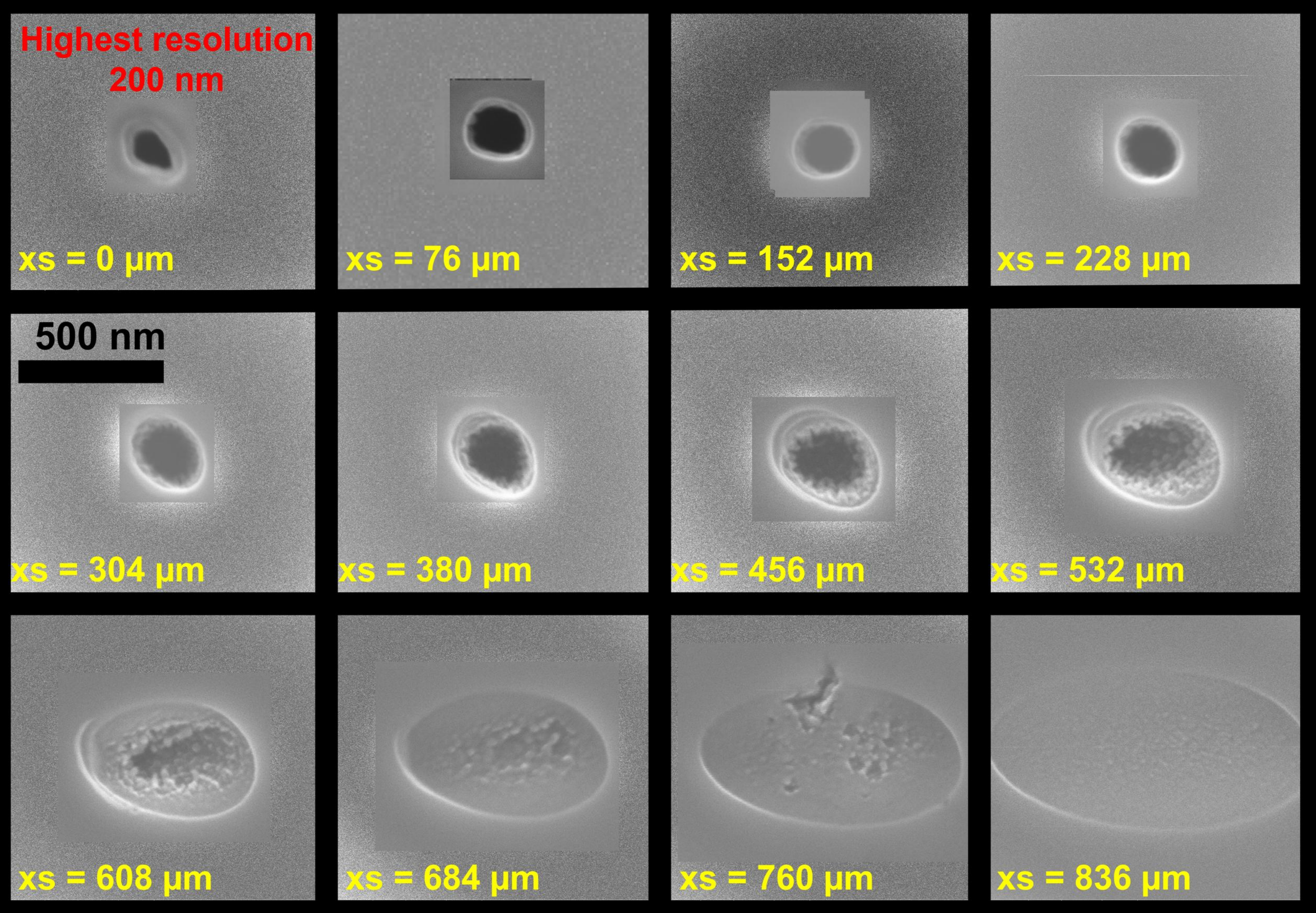}
  \caption{Effect of nozzle-to-beam distance $x_s$ on etching pattern dimensions. SEM images of etched patterns at different $x_s$ positions. Etching conditions: spot scanning, 5\,keV electron energy, 0.935\,pA current, $10^{-4}$\,mbar air, 2\,min exposure. \textbf{Increasing $x_s$ leads to larger, more elongated, and shallower patterns, reflecting the decrease in local gas flux from deeper, localized etching near the nozzle to more diffuse etching at larger distances.}}
  \label{fig:Distance_Dependence_SEM}
\end{figure}

Since the operating pressure in the SEM chamber during EBIE is maintained in the range of \(10^{-6}\) to \(2 \times 10^{-4}\) mbar---several orders of magnitude lower than atmospheric pressure---the gas dynamics inside the chamber are governed by free molecular flow. Under these conditions, the total gas throughput remains effectively constant for a fixed chamber pressure and is balanced by the pump-out rate of the vacuum system. However, the local gas flux emitted from the injection nozzle, referred to the \textit{impinging flux}, exhibits a non-uniform spatial distribution across the sample surface due to geometric constraints and directional flow from the nozzle. This spatial variation in molecular flux plays a important role in defining the localized etching behavior during EBIE.

Figure~\ref{fig:Distance_Dependence}a shows the schematic of the nozzle–beam configuration. The influence of the lateral position $x_s$ between the gas nozzle and the electron beam on EBIE etching behavior is presented in Figure~\ref{fig:Distance_Dependence}, and corresponding SEM images of the etched patterns at various $x_s$ positions are shown in Figure~\ref{fig:Distance_Dependence_SEM}.

Etching experiments were conducted by focusing a 5\,keV electron beam at a fixed position on the diamond surface for 2 minutes under an air atmosphere at $10^{-4}$\,mbar. At the initial position, $x_s = 0$, the resulting etched pattern exhibits a circular shape with a diameter of approximately 200\,nm. As $x_s$ increases, the pattern remains circular up to $x_s = 228\,\mu$m. Beyond this point, the shape gradually transitions into an ellipse, with both the length and width increasing, while the etching depth decreases. At $x_s = 836\,\mu$m, the etched pattern presents an elliptical shape with a major-axis length of 1.24\,$\mu$m. The pattern size—defined as the diameter for circular patterns or the major-axis length for elliptical patterns—is plotted as square symbols in Figure~\ref{fig:Distance_Dependence} as a function of $x_s$.

The red star symbols in Figure~\ref{fig:Distance_Dependence}b represent a simulation of the normalized impinging molecular flux $J_i/J_{\mathrm{tot}}$ along the lateral coordinate $x_s$ (at $y_s = 0$), calculated under the assumption of a free molecular flow regime. In this regime, molecular trajectories are not influenced by intermolecular collisions, and the local flux distribution can be derived geometrically. The impinging flux is given by~\cite{utke2012nanofabrication}:

\begin{equation}
  \frac{J_i}{J_{\mathrm{tot}}} = \left( \frac{\sin(\alpha + \Theta(x_s))}{\sin(\alpha)} \right)^2 \cos(\Theta(x_s)) \sin(\alpha + \Theta(x_s))
\end{equation}

where $\alpha$ is the angle between the nozzle and the substrate, and $\Theta(x_s)$ is defined as:

\begin{equation}
  \Theta(x_s) = \arctan\left( \frac{\sin\alpha}{\frac{L_s}{x_s} - \cos\alpha} \right)
\end{equation}

Here, $L_s$ denotes the distance from the nozzle tip to the substrate along the nozzle axis (see Figure~\ref{fig:Distance_Dependence}a). In our experimental configuration, $\alpha = 30^\circ$ and $L_s = 650\,\mu\mathrm{m}$. Under these parameters, the impinging flux reaches a maximum near $x_s = 424\,\mu\mathrm{m}$ and decreases sharply at larger $x_s$ values. Slightly negative values predicted near $x_s \approx L_s$ arise from singularities in the trigonometric expression and can be considered numerically negligible. A Monte Carlo simulation reported in Ref.~\cite{utke2012nanofabrication} yields a similar flux profile.

When the PE beam strikes the diamond surface, SEs are emitted isotropically. As previously discussed in Section~\ref{sect:Energy_Dependence}, these SEs interact with gas molecules near the surface, initiating dissociation events that produce reactive radicals. The mean free path of SEs depends on the local gas molecule density~\cite{utke2012nanofabrication}. At short nozzle-to-beam distances—where the impinging flux is high—the mean free path is reduced, and SE–gas scattering is confined near the beam center. This results in a high local concentration of reactive species and a tightly localized chemisorption region. Consequently, the etched features formed near the nozzle ($x_s \leq 228\,\mu\text{m}$) tend to be narrow and deep.

At larger $x_s$ values, the impinging flux—and thus the local gas density—decreases, increasing the SE mean free path. The resulting scattering volume expands, leading to a broader chemisorption zone and more diffuse radical-surface interactions. This produces wider, shallower etching profiles. These observations suggest that the etching characteristics are governed by the local gas density. As such, the EBIE process is highly sensitive to the relative spatial position between the gas nozzle and the irradiated region, and the highest etching resolution - 200\,nm is achieved when the beam is placed near the gas nozzle.

\subsection{Influence of electron current density}
\label{sect:ElectronDensity}

\begin{figure}[!t]
  \centering
  \includegraphics[width=\columnwidth]{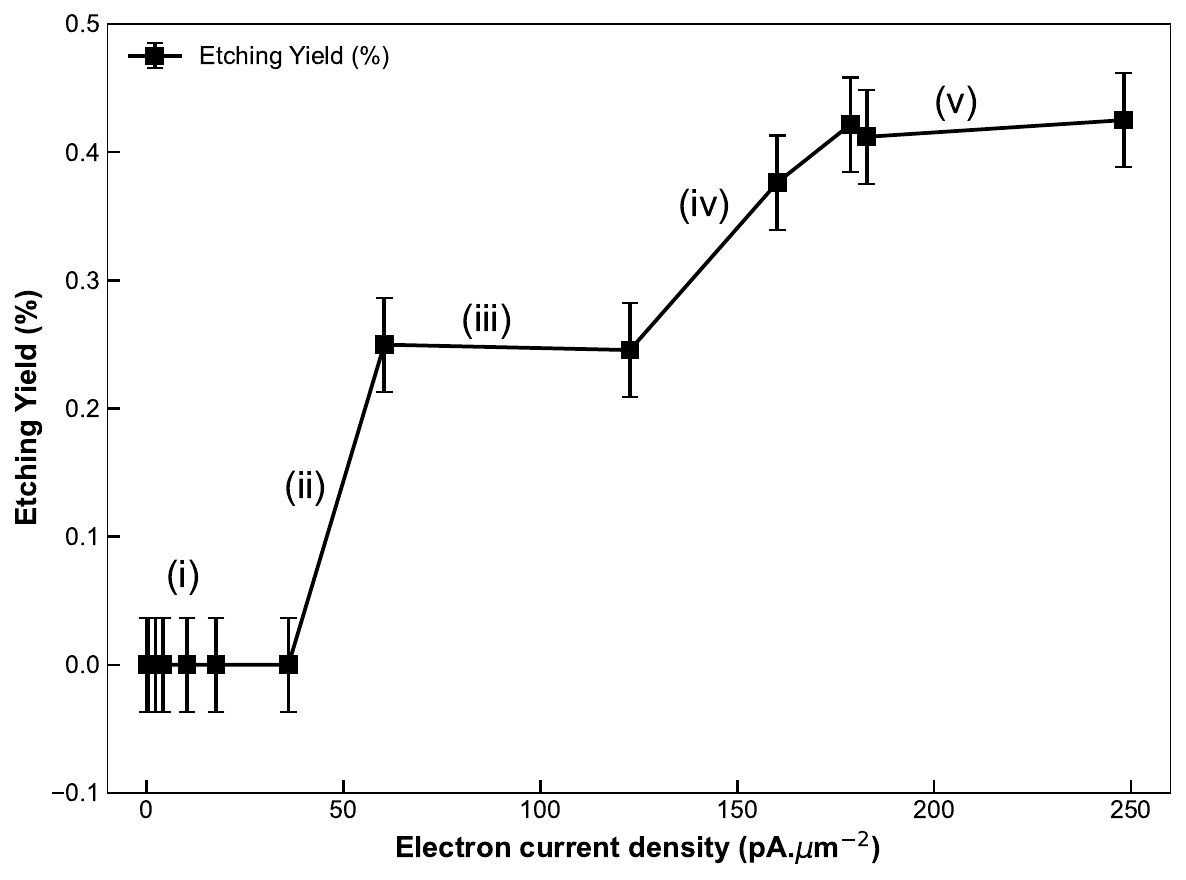}
  \caption{Etching yield as a function of primary electron current density. Etching conditions: 2.5\,$\mu$m\,$\times$\,2.5\,$\mu$m scanned area, 3\,keV electron energy, $10^{-4}$\,mbar air, 5\,min exposure. Error bars reflect surface roughness variations of unetched regions. \textbf{The etching yield exhibits two plateau regions, indicating a transition between desorption-limited regimes as the current density increases.}}
  \label{fig:ElectronDensity_Dependence}
\end{figure}

Figure~\ref{fig:ElectronDensity_Dependence} shows the dependence of the etching yield on the primary electron PE current density, defined as the electron current per unit exposure area, under fixed conditions of 3\,keV beam energy, $10^{-4}$\,mbar air pressure, and a 5-minute exposure. The data exhibit two distinct plateau regions—labeled (iii) and (v)—which indicate changes in the dominant surface desorption mechanisms as the current density increases.

The desorption energies of volatile species formed on diamond surfaces vary considerably with their chemical composition. For instance, CO desorption from single-crystalline diamond surfaces requires approximately 1.8–1.9 eV~\cite{enriquez2021oxidative}, whereas breaking the C–CN bond demands 4.6–5.7 eV~\cite{kosar2019benchmark}. In comparison, the thermal energy available at room temperature is only about 25 meV, which is far too low to thermally activate the desorption of these species. However, the PE beam provides the necessary energy either through direct electron impact or via localized beam-induced heating at the irradiated region. This localized heating is directly governed by the instantaneous electron current density: higher current densities lead to increased local temperature rise, thereby facilitating the desorption of surface chemisorption products with higher binding energies. Consequently, the observed evolution of the etching yield with current density reflects a transition in the dominant desorption mechanism, from species with lower desorption energy (e.g., CO) at moderate current densities to those with higher desorption thresholds (e.g., CN) at higher current densities.

The trends observed in Figure~\ref{fig:ElectronDensity_Dependence} can be interpreted as follows:

\begin{itemize}
  \item[(i)] \textbf{Electron current density $<$ 36\,pA/$\mu$m$^{-2}$:} Insufficient energy is delivered to enable desorption of either CO or CN species. No measurable etching occurs.

  \item[(ii)] \textbf{36–60\,pA/$\mu$m$^{-2}$:} The available energy surpasses the desorption threshold for CO, but remains below that of CN. Etching proceeds via CO desorption only.

  \item[(iii)] \textbf{60–122\,pA/$\mu$m$^{-2}$:} Desorption of CO species dominates and reaches saturation under the existing gas flux. This regime corresponds to the first plateau in etching yield.

  \item[(iv)] \textbf{122–179\,pA/$\mu$m$^{-2}$:} Energy becomes sufficient to initiate CN desorption in addition to CO. As both species contribute to etching, the yield increases beyond the first plateau.

  \item[(v)] \textbf{$>$ 179\,pA/$\mu$m$^{-2}$:} Full desorption of both CO and CN species is achieved. The etching yield saturates again, forming the second plateau.
\end{itemize}

\subsection{Influence of gas pressure}

\begin{figure}[!t]
  \centering
  \includegraphics[width=\columnwidth]{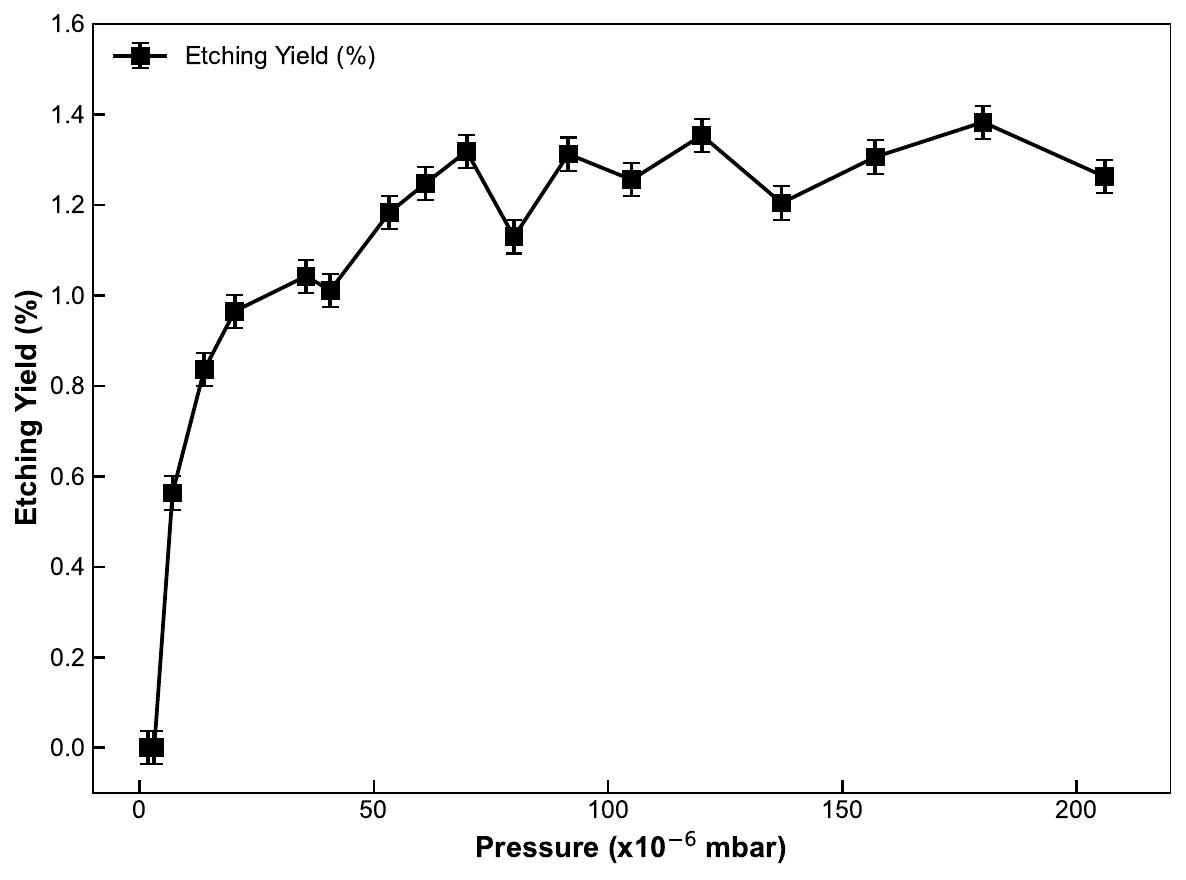}
  \caption{Etching yield as a function of SEM chamber pressure during air injection. Etching conditions: 2.5\,$\mu$m\,$\times$\,2.5\,$\mu$m scanned area, 3\,keV electron energy, electron current density of 183\,pA\,$\mu$m$^{-2}$, 5\,min exposure. Error bars reflect standard deviation from surface roughness of unetched regions. \textbf{The yield increases with pressure up to $\sim5 \times 10^{-5}$\,mbar before saturating, indicating a transition from molecule-limited to electron-limited regimes.}}
  \label{fig:Pressure_Dependence}
\end{figure}

Figure~\ref{fig:Pressure_Dependence} presents the dependence of etching yield on SEM chamber pressure over the range of $10^{-6}$ to $2 \times 10^{-4}$~mbar during air injection. The yield increases sharply with pressure up to approximately $5.35 \times 10^{-5}$~mbar, beyond which it saturates and remains constant despite further increases in pressure.

This behavior is attributed to the transition between two distinct regimes governed by the availability of gas molecules and the electron-induced dissociation rate. At low pressures, the process is limited by the molecular flux to the surface (molecule-limited regime), while at higher pressures, the electron current becomes the limiting factor (electron-limited regime).

\begin{itemize}
  \item \textbf{Molecule-limited regime} ($P < 5.35 \times 10^{-5}$~mbar):
    At low pressures, the flux of gas molecules reaching the surface is limited. Under fixed electron current density, nearly all adsorbed gas molecules are dissociated by secondary electrons. As the pressure increases, the molecular flux also increases, resulting in a higher etching yield.

  \item \textbf{Electron-limited regime} ($P > 5.35 \times 10^{-5}$~mbar):
    At higher pressures, the surface becomes saturated with gas molecules. The dissociation rate is then limited by the availability of electrons. All incident electrons are utilized in the dissociation process, and further increases in gas flux do not enhance the yield. As a result, the etching yield remains constant at fixed electron current density.
\end{itemize}

\subsection{Influence of Dwell Time}

\begin{figure}[!t]
  \centering
  \subfloat[]{\includegraphics[width=0.36\linewidth,height=0.24\textheight,keepaspectratio]{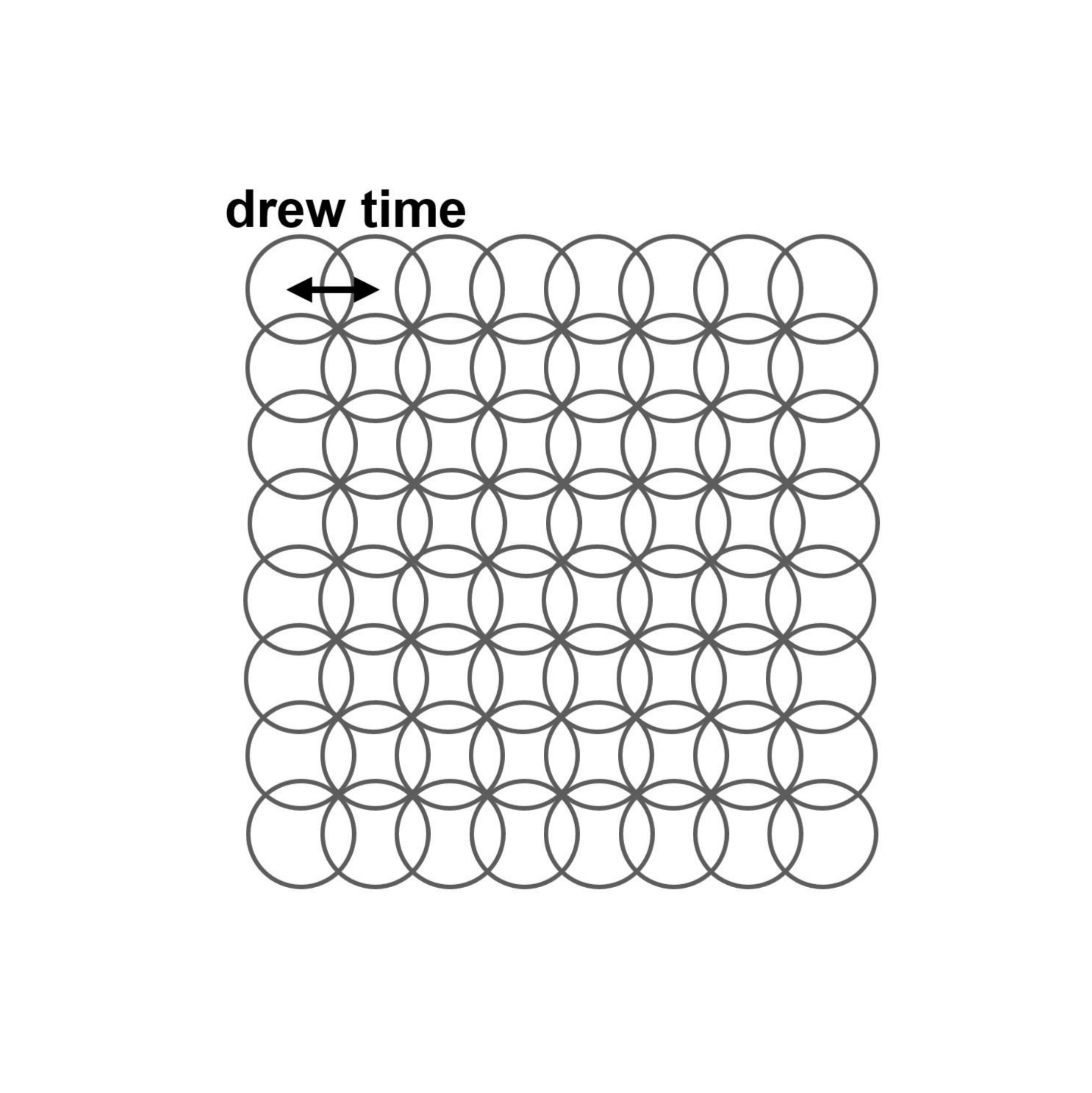}}\hfill
  \subfloat[]{\includegraphics[width=0.62\linewidth,height=0.24\textheight,keepaspectratio]{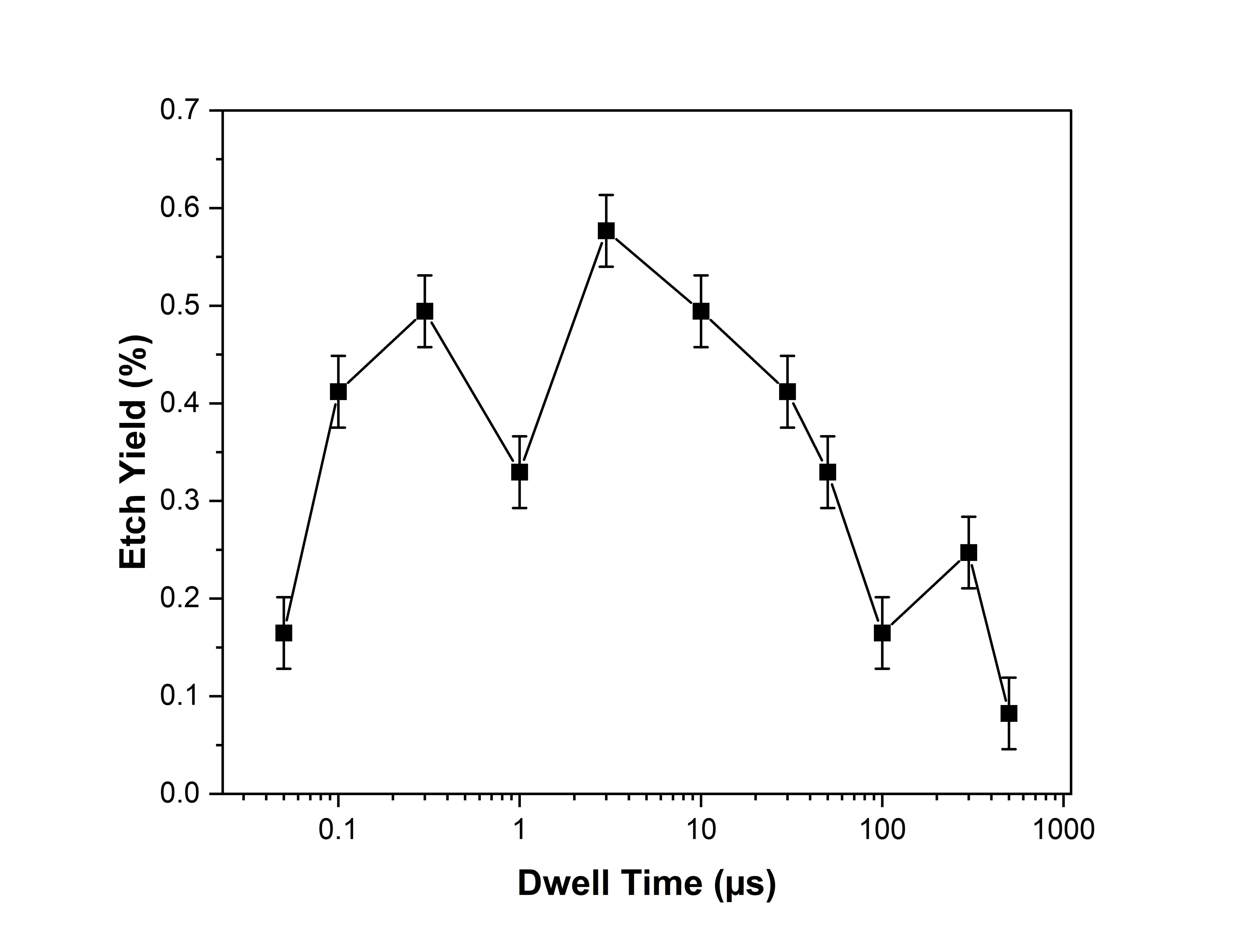}}
  \caption{(a) Schematic representation of scanning dwell time and (b) influence of dwell time on etching yield. Etching conditions: 2.5\,$\mu$m\,$\times$\,2.5\,$\mu$m scanned area, 3\,keV electron energy, $10^{-4}$\,mbar air pressure, 183\,pA\,$\mu$m$^{-2}$ electron current density, 5\,min exposure. Error bars reflect standard deviation from surface roughness of unetched regions. \textbf{The etching yield exhibits a non-monotonic dependence on dwell time, with a maximum yield observed at 3\,$\mu$s due to the optimal balance between electron flux and gas molecule availability}.}
  \label{fig:DwellTime_Impact}
\end{figure}

The dwell time, Figure~\ref{fig:DwellTime_Impact}a, defined as the duration the electron beam remains on each pixel, significantly influences the etching yield. Figure~\ref{fig:DwellTime_Impact}b shows that the etching yield varies non-monotonically with dwell time, with a maximum yield of 0.6\% at 3~$\mu$s.

Notably, longer dwell times correspond to higher instantaneous electron current density per unit area. At extremely short dwell times (50--100~ns), the electron flux is insufficient to generate a significant quantity of reactive radicals, resulting in electron-limited etching. In contrast, at extremely long dwell times (300--500~$\mu$s), excessive electron dose leads to saturation of re-emitted SEs beyond the available gas flux, reducing etch efficiency and placing the process in the molecule-limited regime. The maximum etching yield at intermediate dwell times arises from an optimal balance between electron flux and gas molecule availability.

\subsection{Etching time and Anisotropy}
\label{sect:Time_Dependence}

\begin{figure}[!t]
  \centering
  \includegraphics[width=\columnwidth]{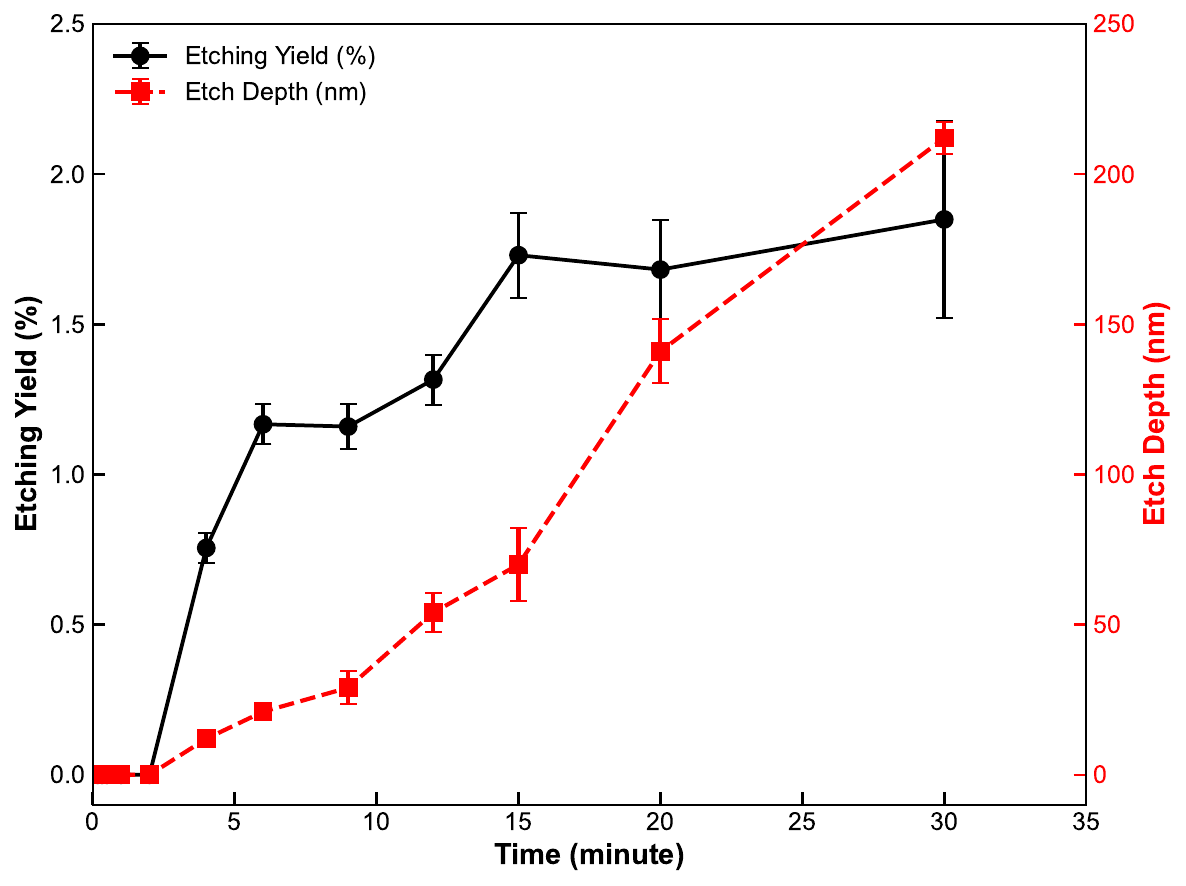}
  \caption{Etching yield and depth as a function of electron exposure time. Etching conditions: 2.5\,$\mu$m\,$\times$\,2.5\,$\mu$m scanned area, 3\,keV electron energy, 183\,pA\,$\mu$m$^{-2}$ current, $10^{-4}$\,mbar air. Error bars reflect surface roughness of unetched regions and standard deviations from batch measurements. \textbf{Both etching yield and depth increase with exposure time, with two distinct yield plateaus observed at 6–9\,min and 15–30\,min.}}
  \label{fig:Time_Dependence}
\end{figure}

\begin{figure}[!t]
  \centering
  \includegraphics[width=\columnwidth]{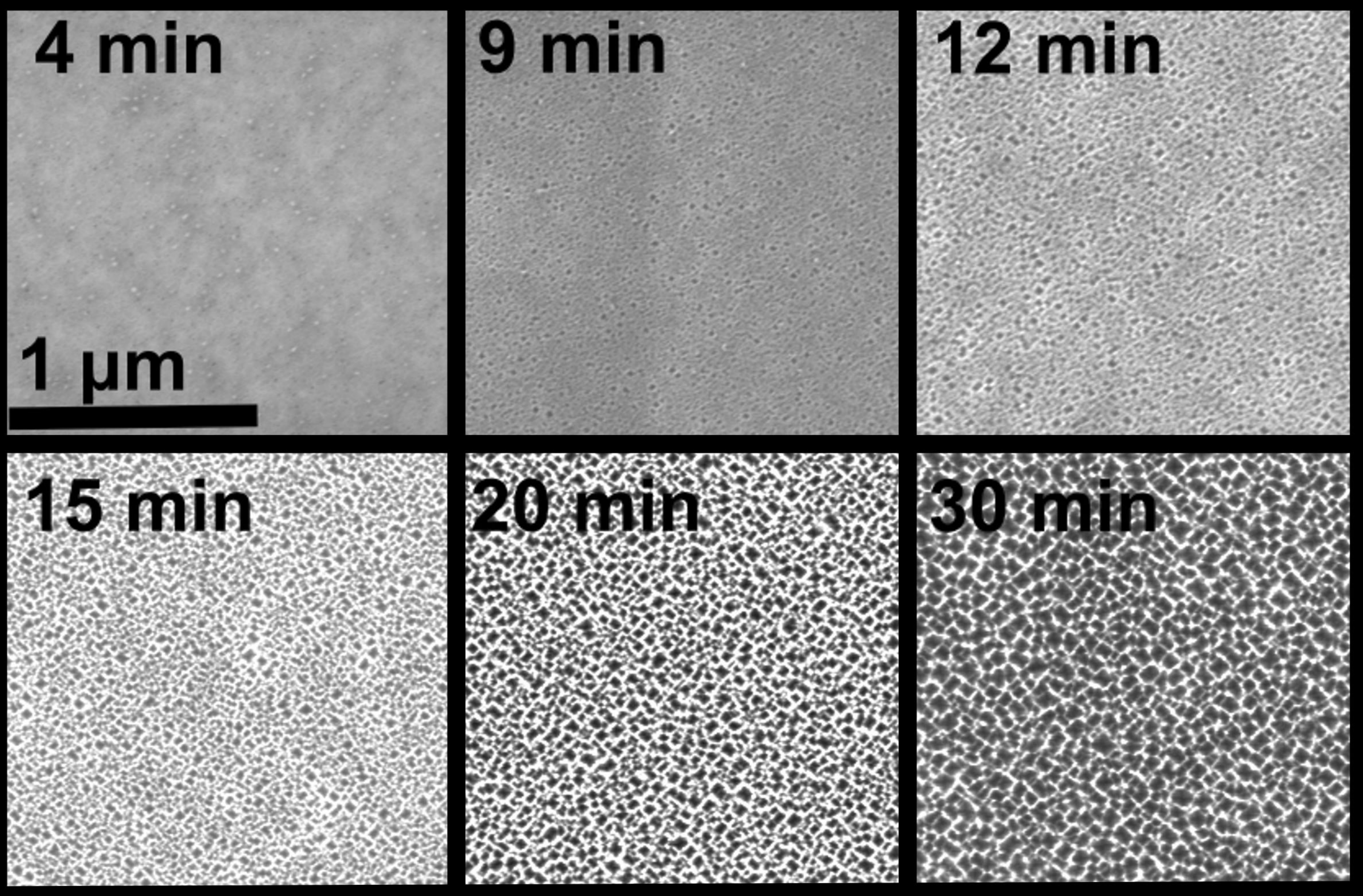}
  \caption{SEM images of the etched surface at different exposure times. Etching conditions: 2.5\,$\mu$m\,$\times$\,2.5\,$\mu$m scanned area, 3\,keV electron energy, 183\,pA\,$\mu$m$^{-2}$ current, $10^{-4}$\,mbar air. \textbf{Anisotropic inverse-pyramid etch pits appear from 9\,min onward, with both the density and size of pits increasing progressively with exposure time.}}
  \label{fig:Time_Dependence_SEM}
\end{figure}

Figure~\ref{fig:Time_Dependence} shows the evolution of etching yield and depth as a function of electron exposure time. A maximum etch depth of 212\,nm is achieved after 30 minutes of irradiation.

No etching is observed for exposure times below 2 minutes, which is attributed to the presence of surface contaminants. Electron-induced reactions with these contaminants result in a transient deposition process that inhibits etching. During this phase, the formation of volatile by-products is suppressed, and no discernible surface modification occurs.

Beyond 2 minutes, progressive removal of the contamination layer enables effective interaction between incident electrons and adsorbed gas molecules, initiating etching. The etch depth increases steadily as the diamond surface becomes increasingly reactive.

For exposure times exceeding 4 minutes, significant material removal is observed. The etching yield stabilizes at approximately 1.2\% between 4 and 9 minutes, then gradually increases to around 1.68\% - 1.85\%  at 15–20 minutes. This progressive enhancement correlates with morphological changes in the etched surface.

Figure~\ref{fig:Time_Dependence_SEM} reveals that up to 9 minutes, the surfaces remain relatively smooth. Beyond this point, anisotropic etch pits with truncated pyramidal morphologies begin to emerge at the bottom of the etched regions. Both the size and density of these pits increase with exposure time, resulting in a surface densely covered with pyramidal features after 15–20 minutes, with an estimated density of $10^8$~cm$^{-2}$.

These pits are bounded by (111) sidewalls, as expected for etching on (100)-oriented diamond, where the crystal structure favors cleavage along the energetically stable \{111\} planes. The emergence of these facets significantly increases the effective surface area exposed to primary electrons and reactive species generated by gas dissociation.
Geometrically, the development of inverted (111) pyramids leads to an increase in effective surface area of approximately 74\%, given by:

\begin{equation}
  \text{Surface increase} = \left( \frac{1}{\cos 55^\circ} - 1 \right) \approx 74\%
\end{equation}

where 55$^\circ$ is the angle between the (100) and (111) crystallographic planes in diamond.

This enlargement enhances local collision rates, secondary electron emission, and chemisorption efficiency, which coherently accounts for the observed rise in etching yield from 1.16\% at 9 minutes to 1.85\% at 30 minutes, corresponding to an increase of approximately 60\%.

Once the surface becomes fully covered by these etch pits after approximately 15 minutes, the effective surface area reaches saturation and remains constant. Consequently, the etching yield also stabilizes, as reflected in the plateau observed in Figure~\ref{fig:Time_Dependence} between 15 and 30 minutes.

The formation of (111) facets under air-based EBIE conditions contrasts with the isotropic morphologies typically reported under pure oxygen atmospheres~\cite{bishop2018deterministic}. This anisotropy is attributed to the combined etching actions of both oxygen and nitrogen radicals present in air.

\subsection{Etching Mechanism}

\begin{figure}[!t]
  \centering
  \includegraphics[width=\columnwidth]{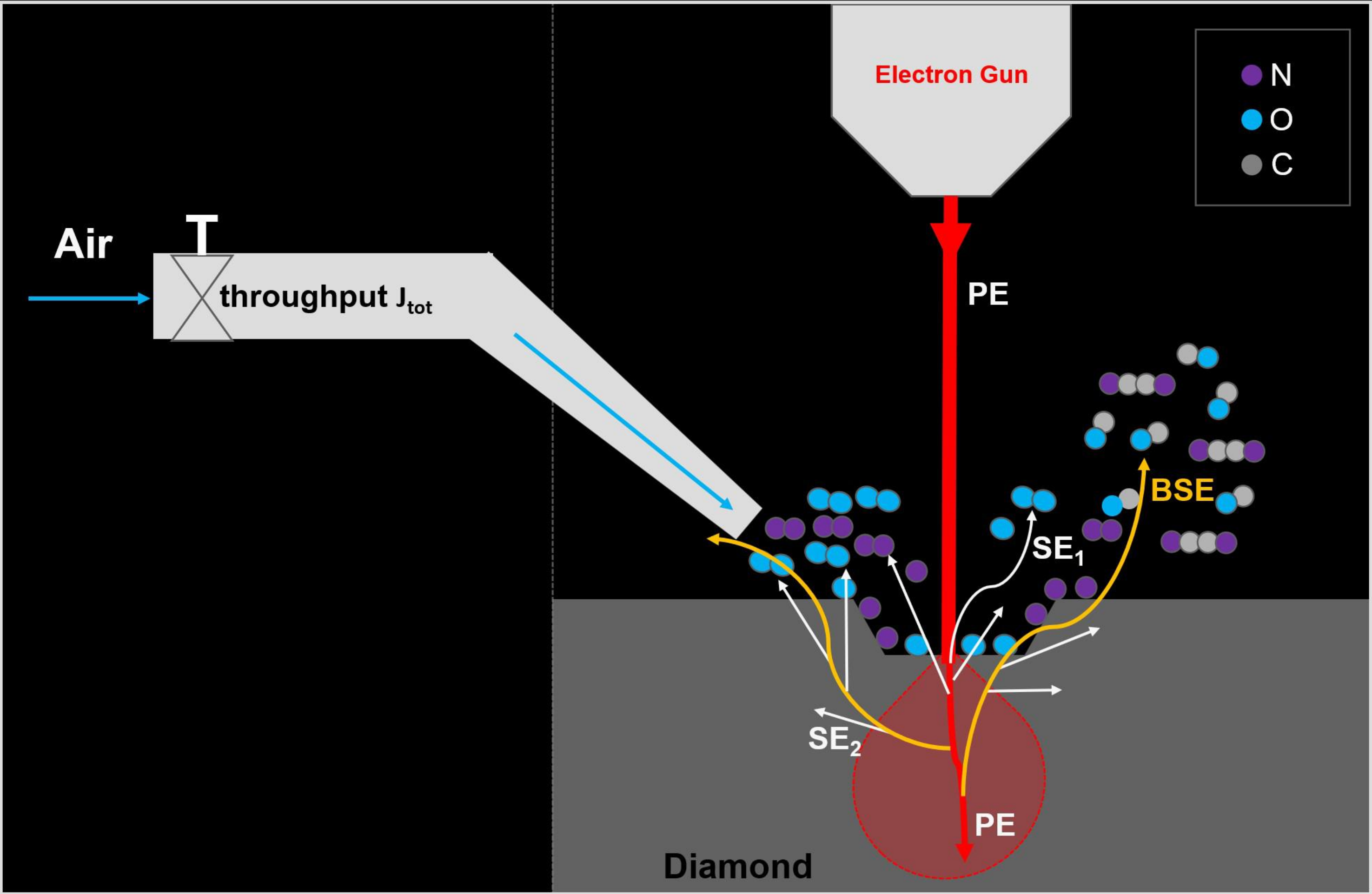}
  \caption{Schematic illustration of the EBIE mechanism under air atmosphere.}
  \label{fig:EBAE_Mechanism}
\end{figure}

The  mechanism of electron beam–induced etching EBIE in air is illustrated in Figure~\ref{fig:EBAE_Mechanism}. Primary electrons, with energies in the  keV range, strike the diamond surface and generate cascades of low-energy secondary electrons, typically around 50~eV. These SEs interact with adsorbed air molecules—primarily O$_2$ and N$_2$—leading to their dissociation into highly reactive oxygen and nitrogen radicals.

These radicals chemisorb onto the diamond surface, forming transient surface complexes that subsequently decompose into volatile by-products such as CO, CO$_2$, and CN. Continued electron irradiation promotes desorption of these volatile species, resulting in the localized removal of carbon atoms from the diamond lattice.

At low gas pressures or with long electron dwell times, the process is \textit{molecule-limited}: the etch rate is restricted by the limited flux of gas species impinging on the surface. In contrast, at higher pressures or with short dwell times, the process becomes \textit{electron-limited}: all SEs are fully utilized for molecular dissociation, and the etch rate scales with the electron current density.

With increasing exposure time, the etching process transitions from isotropic to anisotropic due to the crystallographic anisotropy of diamond. The (100) plane etches more rapidly than the densely packed \{111\} planes, which exhibit higher atomic density and bond energy. Consequently, on (100)-oriented substrates, etch pits evolve into truncated pyramids bounded by four (111) sidewalls.

This anisotropic morphology increases the effective surface area exposed to both incident electrons and dissociated radicals, thereby enhancing the local reaction rate. The observed rise in etching yield at longer exposure durations reflects this enhanced surface reactivity, as discussed in Section~\ref{sect:Time_Dependence}.

Previous studies have reported that EBIE performed under pure oxygen leads to isotropic etching behavior \cite{bishop2018deterministic}. In contrast, the formation of anisotropic features during air-based EBIE suggests a synergistic effect between oxygen and nitrogen radicals. Nitrogen, in particular, may contribute to anisotropic behavior by forming stable surface intermediates or altering local reaction pathways, although its exact mechanistic role remains to be fully elucidated.

\section{Conclusion}

Electron beam–induced etching EBIE in air provides a practical and versatile approach for controlled diamond etching, fully compatible with standard SEM systems. The etching yield reaches its maximum at a primary electron energy of 3~keV, within the typical operating range of most commercial instruments. Optimal etching resolution of 200\,nm is achieved in close proximity to the gas injection nozzle.

At chamber pressures between $10^{-5}$ and $10^{-4}$\,mbar, the etching yield remains stable for an electron current density of approximately 179\,pA/$\mu$m$^{-2}$. Since most SEMs operate with beam currents in the nanoampere range, EBIE enables precise patterning of features with lateral dimensions on the micrometre scale. For larger-area etching, higher beam currents are required.

Etching in air proceeds anisotropically, with the (100) surface orientation preserved at short exposure times and progressive formation of (111)-faceted pits at longer durations. This crystallographically selective behavior enhances the etch yield through increased surface area exposure.

EBIE in air provides a maskless, scalable, and highly controllable platform for the anisotropic etching of diamond, offering a promising solution for the fabrication of advanced photonic, quantum, and electronic devices where preserving surface integrity and avoiding subsurface damage are essential for maintaining coherence, minimizing noise, and ensuring long-term stability.

Furthermore, the fundamental understanding of secondary electron–driven mechanisms established here can be extended to enable precise, damage-free etching of other chemically inert systems, including wide-bandgap semiconductors and emerging 2D materials such as graphene. This approach holds promise for advancing quantum technologies and nanoscale photonic devices where surface quality and dimensional control are critical.

\section*{Acknowledgment}

This work was funded by IDEX ISP 2020-l’initiative d’excellence (Idex) Université Grenoble Alpes.

\bibliographystyle{unsrt}

\bibliography{References}

@article{umezawa2018recent,
  title={Recent advances in diamond power semiconductor devices},
  author={Umezawa, Hitoshi},
  journal={Materials Science in Semiconductor Processing},
  volume={78},
  pages={147--156},
  year={2018},
  publisher={Elsevier}
}

@article{volpe2010high,
  title={High breakdown voltage Schottky diodes synthesized on p-type CVD diamond layer},
  author={Volpe, Pierre-Nicolas and Muret, Pierre and Pernot, Julien and Omns, Franck and Teraji, Tokuyuki and Jomard, François and Planson, Dominique and Brosselard, Pierre and Dheilly, Nicolas and Vergne, Bertrand and others},
  journal={physica status solidi (a)},
  volume={207},
  number={9},
  pages={2088--2092},
  year={2010},
  publisher={Wiley Online Library}
}

@article{mi2020integrated,
  title={Integrated photonic devices in single crystal diamond},
  author={Mi, Sichen and Kiss, Marcell and Graziosi, Teodoro and Quack, Niels},
  journal={Journal of Physics: Photonics},
  volume={2},
  number={4},
  pages={042001},
  year={2020},
  publisher={IOP Publishing}
}

@article{pomorski2013super,
  title={Super-thin single crystal diamond membrane radiation detectors},
  author={Pomorski, Michal and Caylar, Benoit and Bergonzo, Philippe},
  journal={Applied physics letters},
  volume={103},
  number={11},
  year={2013},
  publisher={AIP Publishing}
}

@article{toros2020reactive,
  title={Reactive ion etching of single crystal diamond by inductively coupled plasma: State of the art and catalog of recipes},
  author={Toros, Adrien and Kiss, Marcell Kristof and Graziosi, Teodoro and Mi, Sichen and Berrazouane, R and Naamoun, M and Vukajlovic Plestina, J and Gallo, P and Quack, Niels},
  journal={Diamond and Related Materials},
  volume={109},
  pages={107839},
  year={2020}
}

@article{donato2019diamond,
  title={Diamond power devices: state of the art, modelling, figures of merit and future perspective},
  author={Donato, Nazareno and Rouger, Nicolas and Pernot, Julien and Longobardi, Giorgia and Udrea, Florin},
  journal={Journal of Physics D: Applied Physics},
  volume={53},
  number={9},
  pages={093001},
  year={2019},
  publisher={IOP Publishing}
}

@article{kawabata2004xps,
  title={XPS studies on damage evaluation of single-crystal diamond chips processed with ion beam etching and reactive ion beam assisted chemical etching},
  author={Kawabata, Yusaku and Taniguchi, Jun and Miyamoto, Iwao},
  journal={Diamond and related materials},
  volume={13},
  number={1},
  pages={93--98},
  year={2004},
  publisher={Elsevier}
}

@article{taniguchi1994electron,
  title={Electron beam assisted etching of single crystal diamond chips},
  author={Taniguchi, Jun and Miyamoto, Iwao},
  journal={MRS Online Proceedings Library (OPL)},
  volume={354},
  year={1994},
  publisher={Cambridge University Press}
}

@article{taniguchi1997electron,
  title={Electron beam assisted chemical etching of single-crystal diamond substrates with hydrogen gas},
  author={Taniguchi, Jun Taniguchi Jun and Miyamoto, Iwao Miyamoto Iwao and Ohno, Naoto Ohno Naoto and Kantani, Ken'ichi Kantani Ken'ichi and Komuro, Masanori Komuro Masanori and Hiroshima, Hiroshi Hiroshima Hiroshi},
  journal={Japanese journal of applied physics},
  volume={36},
  number={12S},
  pages={7691},
  year={1997},
  publisher={IOP Publishing}
}

@article{bishop2018deterministic,
  title={Deterministic nanopatterning of diamond using electron beams},
  author={Bishop, James and Fronzi, Marco and Elbadawi, Christopher and Nikam, Vikram and Pritchard, Joshua and Froch, Johannes E and Duong, Ngoc My Hanh and Ford, Michael J and Aharonovich, Igor and Lobo, Charlene J and others},
  journal={ACS nano},
  volume={12},
  number={3},
  pages={2873--2882},
  year={2018},
  publisher={ACS Publications}
}

@article{niitsuma2006nanoprocessing,
  title={Nanoprocessing of diamond using a variable pressure scanning electron microscope},
  author={Niitsuma, Jun-ichi and Yuan, Xiao-li and Koizumi, Satoshi and Sekiguchi, Takashi},
  journal={Japanese journal of applied physics},
  volume={45},
  number={1L},
  pages={L71},
  year={2006},
  publisher={IOP Publishing}
}

@misc{gevantman2006crc,
  title={CRC handbook of chemistry and physics},
  author={Gevantman, LH},
  year={2006},
  publisher={CRC Press}
}

@book{pierson2012handbook,
  title={Handbook of Carbon, Graphite, Diamonds and Fullerenes: Processing, Properties and Applications},
  author={Pierson, H.O.},
  isbn={9780815517399},
  lccn={93029744},
  series={Materials Science and Process Technology Series. Electronic},
  year={2012},
  publisher={Elsevier Science}
}

@book{utke2012nanofabrication,
  title={Nanofabrication Using Focused Ion and Electron Beams: Principles and Applications},
  author={Utke, I. and Moshkalev, S. and Russell, P.},
  isbn={9780199734214},
  lccn={2011028174},
  series={Nanomanufacturing series},
  year={2012},
  publisher={Oxford University Press, USA}
}

@article{ascarelli2001secondary,
  title={Secondary electron emission from diamond: Physical modeling and application to scanning electron microscopy},
  author={Ascarelli, P and Cappelli, E and Pinzari, F and Rossi, MC and Salvatori, S and Merli, PG and Migliori, A},
  journal={Journal of Applied Physics},
  volume={89},
  number={1},
  pages={689--696},
  year={2001},
  publisher={American Institute of Physics}
}

@article{itikawa2006cross,
  title={Cross sections for electron collisions with nitrogen molecules},
  author={Itikawa, Yukikazu},
  journal={Journal of physical and chemical reference data},
  volume={35},
  number={1},
  pages={31--53},
  year={2006},
  publisher={American Institute of Physics for the National Institute of Standards and~…}
}

@article{itikawa2009cross,
  title={Cross sections for electron collisions with oxygen molecules},
  author={Itikawa, Yukikazu},
  journal={Journal of Physical and Chemical Reference Data},
  volume={38},
  number={1},
  pages={1--20},
  year={2009},
  publisher={AIP Publishing}
}

@article{itikawa2005cross,
  title={Cross sections for electron collisions with water molecules},
  author={Itikawa, Yukikazu and Mason, Nigel},
  journal={Journal of Physical and Chemical reference data},
  volume={34},
  number={1},
  pages={1--22},
  year={2005},
  publisher={American Institute of Physics for the National Institute of Standards and~…}
}

@article{enriquez2021oxidative,
  title={Oxidative etching mechanism of the diamond (100) surface},
  author={Enriquez, John Isaac and Muttaqien, Fahdzi and Michiuchi, Masato and Inagaki, Kouji and Geshi, Masaaki and Hamada, Ikutaro and Morikawa, Yoshitada},
  journal={Carbon},
  volume={174},
  pages={36--51},
  year={2021},
  publisher={Elsevier}
}

@article{kosar2019benchmark,
  title={Benchmark DFT studies on C--CN homolytic cleavage and screening the substitution effect on bond dissociation energy},
  author={Kosar, Naveen and Ayub, Khurshid and Gilani, Mazhar Amjad and Mahmood, Tariq},
  journal={Journal of Molecular Modeling},
  volume={25},
  pages={1--13},
  year={2019},
  publisher={Springer}
}

@article{brundle1992encyclopedia,
  title={Encyclopedia of materials characterization: surfaces, interfaces, thin films},
  author={Brundle, C Richard and Evans Jr, Charles A and Wilson, Shaun},
  journal={(No Title)},
  year={1992}
}

@article{portier2023carrier,
  title={Carrier Mobility up to 10 6 cm 2 V- 1 s- 1 Measured in Single-Crystal Diamond by the Time-of-Flight Electron-Beam-Induced-Current Technique},
  author={Portier, A and Donatini, F and Dauvergne, D and Gallin-Martel, M-L and Pernot, J},
  journal={Physical Review Applied},
  volume={20},
  number={2},
  pages={024037},
  year={2023},
  publisher={APS}
}

\end{document}